# Interpreting Convolutional Neural Network Activation Maps with Hand-crafted Radiomics Features on Progression of Pediatric Craniopharyngioma after Irradiation Therapy

Wenjun Yang[1], Chuang Wang[2], Tina Davis[3], Jinsoo Uh[3], Chia-Ho Hua[3], Thomas E. Merchant[3]

1. Department of Radiation oncology, The university of Iowa health and clinics,

    200 Hawkins Dr. Iowa city, IA. 52242

2. Department of Radiation oncology, University of Miami,

    1475 Northwest 12th Avenue, Miami, FL. 33136

3. Department of Radiation oncology, St. Jude Children's research hospital,

    262 Danny Thomas Place, Memphis, TN. 38105

Abstract:

Purpose: Convolutional neural networks (CNNs) are promising in predicting treatment outcome for pediatric craniopharyngioma while the decision mechanisms are difficult to interpret. We compared the activation maps of CNN with hand crafted radiomics features of a densely connected artificial neural network (ANN) to correlate with clinical decisions.

Methods: A cohort of 100 pediatric craniopharyngioma patients were included. Binary tumor progression was classified by an ANN and CNN with input of T1w, T2w, and FLAIR MRI. Hand-crafted radiomic features were calculated from the MRI using the LifeX software and key features were selected by Group lasso regularization, comparing to the activation maps of CNN. We evaluated the radiomics models by accuracy, area under receiver operational curve (AUC), and confusion matrices.

Results: The average accuracy of T1w, T2w, and FLAIR MRI was $0.85 \pm 0.04$, $0.92 \pm 0.03$, and $0.86 \pm 0.08$ (ANOVA, $F = 1.96$, $P = 0.18$) with ANN; $0.83 \pm 0.03$, $0.81 \pm 0.07$, and $0.70 \pm 0.05$ (ANOVA, $F = 10.11$, $P = 0.003$) with CNN. The average AUC of ANN was $0.91 \pm 0.07$, $0.97 \pm 0.02$, and $0.90 \pm 0.06$; $0.86 \pm 0.07$, $0.88 \pm 0.11$, and $0.75 \pm 0.10$ of CNN for the 3 MRI, respectively. The activation maps were correlated with tumor shape, min and max intensity, and texture features.

Conclusions: The tumor progression for pediatric patients with craniopharyngioma achieved promising accuracy with ANN and CNN model. The activation maps extracted from different levels were interpreted with hand-crafted key features of ANN.

**Introduction**

Craniopharyngioma is a benign tumor that most occurs in the suprasellar region of brain, and commonly occurs in children and adolescents (1). The locally aggressive features often result in compression of optic pathways, circle of Willis, and hypothalamic-pituitary axis, with typical symptoms of headache, sleep disorder, and hormone deficiency (2, 3) induced obesity (4, 5). Limited surgery combined with radiation therapy or surgery alone are standard treatments yielding similar outcomes with 5-year survival above 90% and 10-year survival above 80% (6). With the development of irradiation therapy techniques in the past ten years, tumor progression rate has dropped from approximately 10% with photon therapy to 5% with proton irradiation (7, 8). However, current irradiation clinical trials are mostly based on uniform protocols (9, 10), while personalized treatment planning including target dose escalation or organs at risk (OAR) dose control with prognostic information could potentially reduce the progression rate.

Radiomics (11, 12) analyzes medical imaging features to predict treatment outcome or complexities. Deep learning significantly improves prediction accuracy with a high dimensional input. Radiomics was first introduced with hand crafted features calculated with a variety of equations to extract texture and histogram features (13, 14). The curse of dimension problem, which defers to the high dimensional feature space compared to sample size, is usually mitigated by filtering the feature pool (14, 15). However, most pre-selection methods consider the significance of each feature individually while the interplay effect among features is usually neglected (16). We applied an embedded Group lasso regularization in a densely connected artificial neural network (ANN) to select key features by optimizing a sparse distribution of relative weights for each feature.

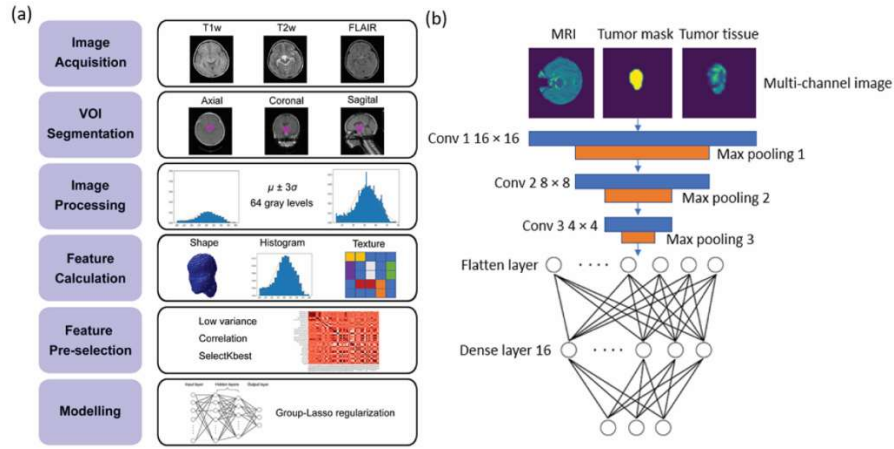

Fig 1. Data processing and modelling workflow. (a) ANN model pipeline. T1w, T2w, and FLAIR MRI were acquired before surgery. The gross tumor volume was segmented by a radiation oncologist. After image preprocessing, different levels of features were calculated within the segmented VOI. A 4-layer densely connected artificial neural network with Group lasso and dropout regularization was trained and evaluated by 5-fold cross validation. (b) CNN model pipeline. Multi-channel image data were created with channel 1 of pre-processed MRI, channel 2 of tumor mask, and channel 3 of tumor tissue by multiplying the previous 2 channels. Feature engineering layers consist of 3 convolutional and max-pooling layers. After a flattening layer and dense layer, tumor progression was predicted with a focal loss function.

Convolutional neural network (CNN) (17) sequentially convolve multiple channel of images with kernels to adaptively learn the filter pattern to predict outcome. The locally invariant convolutional kernels filter the image parallelly to reduce the size of input images (18). CNN is usually more data demanding comparing to densely connected ANN partially due to the enlarged number of channels at each level. Moreover, the adaptively learned activation maps at each layer are difficult to interpret due to the complex patterns (19). As a result, CNN is often overfitting to a training cohort and external validation is usually helpful for generalization.

In this study, we compared CNN activation maps with the key features of ANN selected by the Group lasso regularization based on T1w, T2w, and FLAIR diagnostic MRI for pediatric craniopharyngioma patients treated with limited surgery and irradiation. Tumor progression was determined in the 5-year follow up visits and the binary classification label was the prediction output of radiomics models. The key features selected by ANN with Group lasso regularization were compared to the activation maps extracted from the convolutional layers of CNN, which could facilitate the interpretation of CNN models and integration of mitigation methods. To the best of our knowledge, this is the first study comparing hand-crafted and automatic learned radiomics features with diagnostic MRI on tumor progression for pediatric craniopharyngioma patients.

**Materials and Methods**

*Patient cohort and image acquisition*

This study included a retrospective pediatric patient cohort composed of 100 patients aging from 0 to 24 at treatment, with approval of institutional reviewed board at our hospital. Limited surgery followed by irradiation therapy were prescribed to the patient cohort. Irradiation therapy was delivered with proton for 88 patients, while the remaining 12 patients were treated with photon. Twelve patients after proton and photon therapy, respectively, were diagnosed as progression positive, which resulted in a progression rate of 0.24 in this patient cohort. The irradiation therapy dose was prescribed to 54 Gy (RBE) in 30 fractions with parallel opposed 3D conformal beams. Tumor progression was determined with the follow-up CT and MRI as the re-growth of tumor tissue. The statistics of patient cohort were summarized in Table 1.

Diagnostic MRI with T1w, T2w, and FLAIR sequences were acquired at multiple imaging centers before treatment in our hospital. Based on the varying imaging protocols of external centers, 88, 95, and 83 MRI were available for the 3 sequences, respectively, as summarized in Table 1. The volume of interest (VOI) was manually delineated as the gross tumor volume (GTV) by experienced radiation oncologists.

**TABLE 1. Summary of patients and MRI scanners**

|  | Patient number | |
|---|---|---|
| Photon | 12 (all positive) | |
| Proton | 88 (12 positive) | |
| Scanner manufacture | Magnetic Strength | |
|  | 1.5 T | 3 T |
| Siemens | 33 | 11 |
| GE | 19 | 7 |
| Philips | 11 | 3 |

*MRI pre-processing*

We applied multiple imaging pre-processing steps for ANN to normalize the diagnostic MRI scanned from multiple external imaging centers. MRI resolutions were resampled to 1×1×3 mm (13, 20) for T1w, T2w, and FLAIR sequences. For the ANN model with hand-crafted features, the MRI intensity was normalized to 64 bins with window and level of mean ± 3×standard deviation (21), to match histograms of all images as shown in Fig 1(a). The manually delineated contour of VOI was registered from treatment planning CT to MRI with rigid registration using the LIFEx software (22) (Institut Curie Research Center, Orsay, France).

More complex image preprocessing was implemented for CNN to normalize the variety of imaging protocols. A similar resolution resampling with 1×1×3 mm was registered from CT to MRI. After imaging axis rectification, the contour of VOI was transferred from CT to MRI using the LifeX software. N4 bias correction (23) was then applied to account for magnetic field inhomogeneity; and brain extraction was performed with the FSL library (24, 25) to strip skull. Finally, histogram of all images (26) was matched to redistribute to the same range to normalize the multi-centric acquisition. We input 3-channels of imaging data to the CNN to focus on target: the first channel of the processed MRI; the second channel of the segmented target mask; and the third channel of the segmented VOI by multiplying the previous 2 channels. We kept image slices with tumor more than 100 pixels to maintain sufficient pathological information.

*Radiomics models*

The hand-crafted features were extracted from a four-layer densely connected ANN. First-order macroscopic (e.g. shape, histogram, and conventional index) and second-order microscopic (e.g. gray-level co-occurrence matrix (GLCM) (27), gray-level run-length matrix (GLRM) (28), neighborhood gray-level different matrix (NGLDM) (29), and gray-level zone-length matrix (GLZLM) (30)) features were calculated within the segmented VOI using LifeX. After a low variance filter, a set of 66 features were input into the ANN model. We balanced the number of patients for both progression positive and negative groups using the synthetic minority oversampling technique (SMOTE) (31) to augment the number of patients in the positive group. Dropout (32) and Group-lasso regularizations (33, 34) were applied to mitigate overfitting and select key features: the relative weights (the L2 norm of the column vector in the weight matrix between the input and the first hidden layer) of input features were sparsely distributed by the Group-lasso regularization, and key features were selected by ranking the sparse relative weights.

The CNN models were constructed with 3 convolutional layers, each paired with a maximum pooling layer for feature extraction, and two densely connected layers to predict tumor progression. The kernel size of convolutional layers was 16, 8, and 4, while the densely connected layers contained 16 and 2 neurons. The structure of CNN was tuned by trial-and errors, without explicit hyperparameter tuning. Data augmentation including image flipping, random shift of 10%, and random rotation of 40 degrees were used to augment the size of progression positive group. Focal loss (35) instead of categorical cross entropy function (36) was selected to exponentially emphasize the relative weights of positive group to enhance the true positive rate of classifiers.

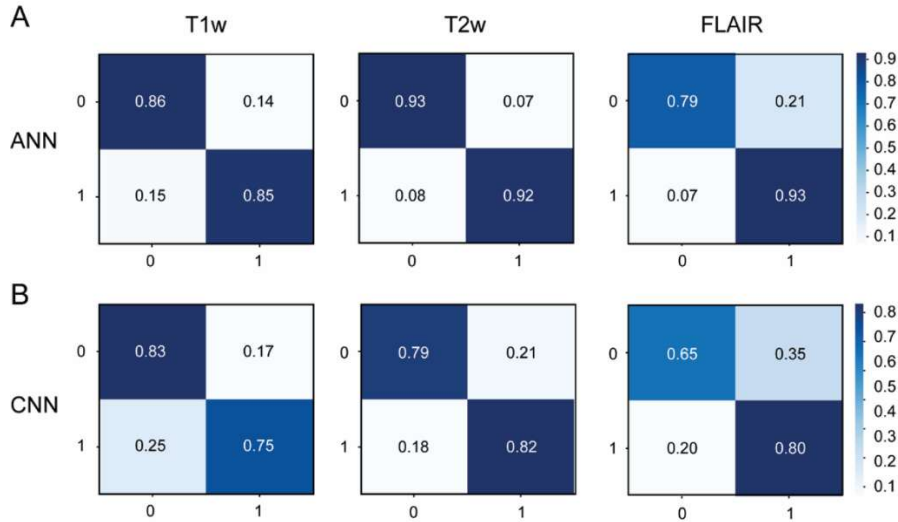

Fig 2. Average confusion matrix of T1w, T2w, and FLAIR MRI. (A) Confusion matrices of ANN averaged over the 5-fold cross validation. (B) Confusion matrices of CNN over the top 5 performance models selected from 10 random experiments.

*Metrics*

Separated ANN and CNN models for T1w, T2w, and FLAIR MRI were developed based on the imaging availability. The entire patient cohort was divided into 80% training and 20% testing set. The ANN models were tested with a 5-fold cross validation, with metrics of average accuracy, AUC, and confusion matrices. The performance of CNN was evaluated from experiments with random 80-20% splitting of the dataset by 10 times. The top 5 performing CNN models were evaluated with the average accuracy, AUC, and confusion matrices. The complete results of the 10 random trials with CNN can be found in the supplementary information.

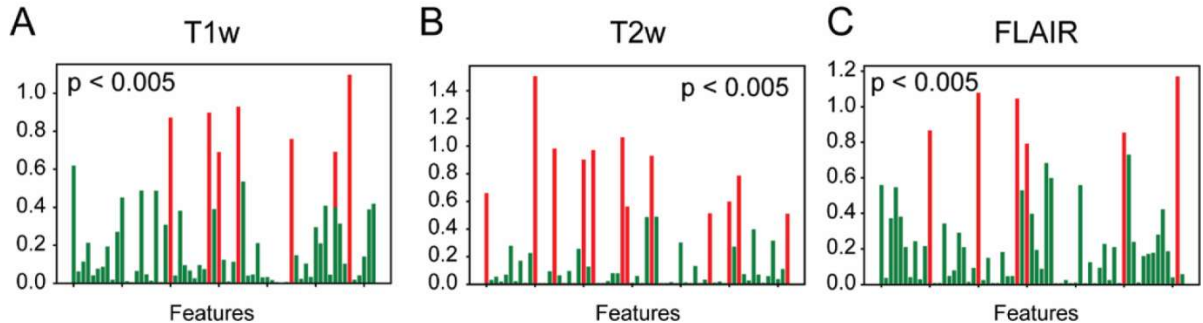

Fig 3. The weights of the features selected by the ANN. The sparsity is forced by the Group lasso regularization. The final features are selected with a *t*-test ($P < .005$) as indicated by the red bars.

**Results**

The average accuracy of ANN models over the 5-fold cross-validation was $0.85 \pm 0.04$, $0.92 \pm 0.03$, and $0.86 \pm 0.08$ (ANOVA, $F = 1.96$, $P = 0.18$) for T1w, T2w, and FLAIR MRI, respectively. The accuracy of ANN models was statistically similar, with the null hypothesis that the mean accuracies were equal. The average AUC was $0.91 \pm 0.07$, $0.97 \pm 0.02$, and $0.90 \pm 0.06$ for ANN, respectively for the 3 MRI sequences. Confusion matrices with principal components greater than 0.8 were observed for 3 sequences as shown in Fig. 2(a). Key features were selected by a 2-sample *t*-test ($P < .005$) with the testing hypothesis that the relative weight of each feature was higher than the average weight of all features over the 5-fold cross-validation, as summarized in Table 2 and illustrated in Fig. 3. The number of key features for the 3 sequences were 7, 12, and 6, respectively.

**TABLE 2. Final features of T1w, T2w, and FLAIR MRI selected by ANN models**

| MRI Sequences | Final Features |
| --- | --- |
| T1W | min, SHAPE_Sphericity, SHAPE_Compacity, GLCM_Contrast (Variance), GLRLM_SRLGE, GLRLM_LRLGE, GLZLM_SZHGE |
| T2W | min, Q1, Skewness, ExcessKurtosis, SHAPE_Sphericity, SHAPE_Surface(mm2), GLCM_Contrast (Variance), GLCM_Correlation, NGLDM_Coarseness, NGLDM_Busyness, GLZLM_LGZE |
| FLAIR | min, SHAPE_Sphericity, GLCM_Contrast (Variance), GLCM_Correlation, NGLDM_Coarseness, GLZLM_ZLNU |

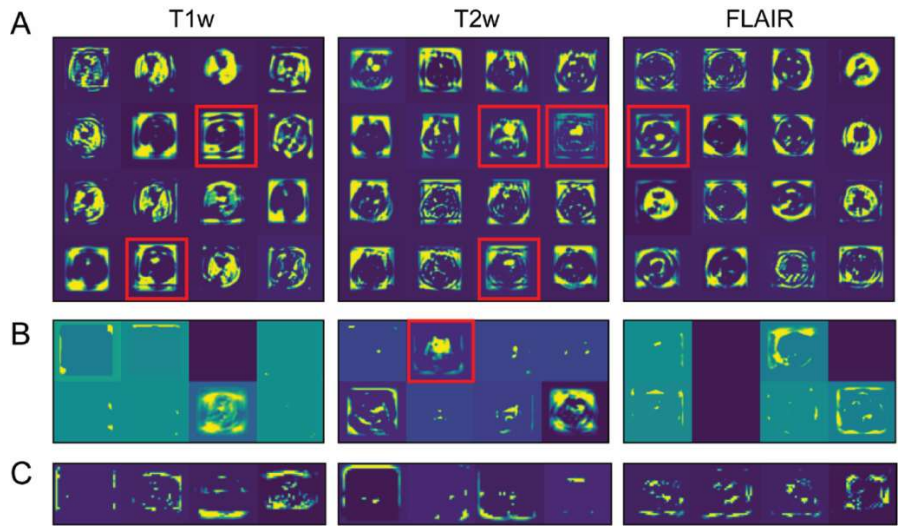

Fig 4. Feature maps extracted from the convolutional layers of CNN for T1w, T2w, and FLAIR MRI. (A) Feature maps of the first convolutional layer of CNN, (B) Features extracted from the second convolutional layer, and (C) Features extracted from the third convolutional layer of CNN. The feature maps with tumor region focus are marked by the red boxes.

The average accuracy of the CNN models over the top 5 random experiments was 0.83 ± 0.03, 0.81 ± 0.07, and 0.70 ± 0.05 (ANOVA, F = 10.11, P = 0.003), which indicated a significantly higher accuracy with T1w and T2w when compared to FLAIR sequence. The average AUC of CNN was 0.86 ± 0.07, 0.88 ± 0.11, and 0.75 ± 0.10 for the 3 sequences, respectively. Lower principal components approximately above 0.7 were obtained with CNN as shown in Fig. 2(b). Feature maps extracted from the convolutional layers of CNN were shown in Figure 4: activation maps became more abstract at deeper levels, and a few features are directly related to the segmented VOI as indicated by the bounding boxes.

**Discussion**

CNN further empowers radiomics by introducing automatic feature learning layers instead of hand-crafted features. However, CNN usually demands a larger dataset due to the expanded parameter space. Moreover, the features could be specific to the data cohort and hard to interpret, which results in a black-box of decision process. A comparison of hand-crafted features and adaptively learned activation maps would benefit interpreting the decision process of CNN.

The performance of ANN with hand-crafted features was superior when compared with the CNN models in terms of accuracy, AUC, and principal component of confusion matrix, which could be caused by the relatively small dataset available for craniopharyngioma that occurs 100-200 cases (37) per year in the U.S. Training a CNN with such a small cohort is subject to overfitting. We tested transfer learning with a pre-trained ResNet-18 (38) on the ImageNet database. The prediction results showed extremely overfit with approximately 0 true positive rates. The capacity of the ResNet-18 is overwhelming for our data cohort. As a result, we developed a compact CNN with focal loss function and balanced data generator to account for the limited data size and unbalanced group distribution. Even with these methods, the CNN models showed stronger dependence on training and testing dataset. The metrics of CNN were unstable during a 5-fold cross-validation when compared to ANN. Thus, we conducted 10 random splitting of training and testing data set, and the top 5 performance models were selected. The metrics of 5-fold cross validation and the 10 random experiments are listed in the supplementary material. Due to the relatively lower dimensions of input features, ANN models showed less overfitting and the performance was relatively stable in the 5-fold cross-validation.

Group-lasso regularization selected key features with interplay effect instead of filtering the features pool sequentially. As summarized in Table 2, minimum intensity, SHAPE_Sphericity, and GLCM_Contrast (Variance) were the common features observed in all the 3 sequences. Ma, et. al. (39) found that tumor shape were highly correlated with invasiveness of adamantinomatous craniopharyngioma, which was consistent with our study that tumor invasiveness could be a main factor in tumor progression. As a comparison, no shape features were selected as key features in our previous radiomics study on post-treatment obesity (40). Intensity histogram and texture features were observed in radiomics studies on classifying benign tumor types (pituitary adenoma, craniopharyngioma, or meningioma) (41-43) or subtypes of craniopharyngioma. Similarly, we found GLCM, GLRLM, and NGLDM features predictive in progression. The minimum intensity represents the difference from the mean (i.e., the mean $-3 \times$ the standard deviation), which represented histogram heterogeneity, was also reported as a key feature in distinguishing tumor progression from radionecrosis after stereotactic radiosurgery (44).

Feature maps extracted from the convolutional layers of CNN are interpretable by the key features of ANN: shape features of brain were widely observed in the first convolutional layer of CNN as shown in Figure 4 (a). The correlation of peripheral brain shape and tumor progression in several CNN channels verified our findings with the ANN models. The second convolutional layer of CNN consists of abstract features with sparse spots which are related to the min or max value of hand-crafted features for ANN. Note that the tumor shape is included in the $2^{nd}$ layer of T2w MRI as marked by the red box. However, some channels eg. channel 1 of the $2^{nd}$ layer of T1w and 3 channels on the same level of FLAIR MRI are empty as shown in Figure 4 (b). These empty channels partially explained the lower performance of T1w and FLAIR when compared to T2w MRI for the CNN models. The feature maps of the $3^{rd}$ convolutional layers were more abstract while they could be related to the texture features of ANN.

Lack of external validation is the main limitation of this study. A relatively low occurrence of craniopharyngioma limits the size of patient cohort and external validation. The performance of ANN and CNN models was also limited by image quality caused by the varying imaging protocols in the external diagnostic imaging centers. These MRI were commonly scanned with a fast-imaging sequence with relatively low image quality, which could compromise the performance of radiomics models. Moreover, the patient cohort was accumulated in the past 30 years, and thus the image quality also varied as the development of MRI techniques. In addition to the image quality and overfitting, the lack of tumor focus further compromised performance for CNN. As shown in the red box of Figure 4, only several feature maps focused on the target, which was highly pathologically related. However, other activation maps without target information were hard to interpret for clinical decisions. Moreover, the relatively low progression rate of around 10% in general population resulted in highly unbalanced dataset, which made the model training even more difficult.

The limitation of data size could be mitigated by external validation, and the complex contour registration could be mitigated by automatic contouring method (45). Image pre-processing and normalization used in this study has effectively reduced the variation caused by the multi-centric MRI with relatively low image quality. To enhance the tumor-focus of CNN, we used multiple channels of imaging and tumor masks to emphasize the target. Vision transformer (46) with self-attention could adaptively focus the pathological related region on the attention maps, which could be the next model of our research.

**Conclusion**

Tumor progression could be estimated with deep-learning based radiomics for pediatric patients with craniopharyngioma. ANN models are suitable for a relatively small data cohort while CNN models are more data demanding. The feature maps of shallow convolutional layers are shape patterns of different regions of brain, and multi-channel MRI with tumor mask could improve the focus on clinical-relevant area. Texture features are observed in deeper layers of CNN, usually without explicit description due to the adaptive learning. The key features selected by ANN models provided insight and reasonable interpretation of the activation maps of CNN. The predictive tumor progression potentially contributes to personalized radiation treatment planning in terms of dose prescription, escalation, coverage, and distribution.

**Acknowledgments**

This study was supported by ALSAC. We extend our sincere appreciation to Keith A. Laycock for his detailed scientific editing, Fang Xie for his support in data collection, and Tina Zhang for her great effort in figure editing.